\documentstyle[sprocl]{article}
\input epsf.tex
\def\Journal#1#2#3#4{{#1} {\bf #2}, #3 (#4)}
\def\ApJ{\em Ap.~J.}
\def\ApJL{\em Ap.~J.~Lett.}
\def\MNRAS{\em Mon.~Not.~Roy.~Ast.~Soc.}
\def\PRL{\em Phys. Rev. Lett.}
\def\onu{\Omega_\nu}
\def\oc{\Omega_c}
\def\om{\Omega_{\rm m}}
\def\ob{\Omega_{\rm b}}
\def\ov{\Omega_\Lambda}
\def\kfs{k_{\rm fs}}
\def\Dnl{\Delta_{\rm nl}}
\def\Dl{\Delta_{\rm l}}
\def\go{\mathrel{\raise.3ex\hbox{$>$}\mkern-14mu
\lower0.6ex\hbox{$\sim$}}}
\def\lo{\mathrel{\raise.3ex\hbox{$<$}\mkern-14mu
\lower0.6ex\hbox{$\sim$}}} \def\onu{\Omega_\nu}

\begin{document}
\title{NEUTRINOS AND DARK MATTER}
\author{CHUNG--PEI MA}
\address{Department of Physics and Astronomy, University of Pennsylvania\\
Philadelphia, PA 19104\\E-mail: cpma@strad.physics.upenn.edu}

\maketitle

\abstracts{ In these lectures I highlight some key features of massive
neutrinos in the context of cosmology.  I first review the thermal
history and the free-streaming kinematics of the uniform cosmic
background neutrinos.  I then describe how fluctuations in the phase
space distributions of neutrinos and other particles arise and evolve
after neutrino decoupling according to the linear perturbation theory
of gravitational instability.  The different clustering properties of
massive neutrinos (aka hot dark matter) and cold dark matter are
contrasted.  The last part discusses the nonlinear stage of
gravitational clustering and highlights the effects of massive
neutrinos on the formation of cosmological structure.}

\section{Neutrino Masses}
These lectures discuss how the universe serves as a learning ground
for massive neutrinos.  Before doing so, let us briefly review some
experimental measurements of neutrino masses.

Upper bounds on neutrino masses from kinematic measurements in
laboratories continue to improve.~\cite{mass} For the $\tau$-neutrino,
$m_{\nu_\tau}< 18.2$ MeV from the decay channel $\tau \rightarrow 5\pi
+ \nu_\tau$.  For the $\mu$-neutrino, $m_{\nu_\mu}< 170$ keV from
two-body pion decay.  For the electron neutrino, the quantity
$m_{\nu_e}^2$ is measured in tritium beta decay by fitting the shape
of the energy spectrum near the endpoint.  Experiments thus far have
yielded nonphysical negative values for $m_{\nu_e}^2$, indicating
unexplained systematic effects in the measurements.  A conservative
upper bound is put at $m_{\nu_e}\approx 15$ eV.  The spread in arrival
times of neutrinos from supernova explosions provides an independent
way to constrain the mass of the electron neutrino.  Various limits
have been reported for SN 1987A; a conservative estimate is
$m_{\nu_e}< 23$ eV.~\cite{mass,mann}

\section{Properties of Cosmic Background Neutrinos}

\subsection{Temperature and Density}

For a brief 1 second after the big bang, neutrinos enjoy being part of
the thermal bath composed of photons, electrons, protons, neutrons,
and the associated anti-particles (after the quark-hadron era).  The
weak interactions at this early time are rapid enough to keep these
particles in thermal equilibrium at a single temperature $T$.  After 1
second, when $T$ drops below about 1 MeV, however, the neutrino
interaction rate becomes slower than the Hubble expansion, and
neutrinos become effectively collisionless and freely-streaming
particles whose trajectories are determined by the geodesic equations.
This event is commonly referred to as ``neutrino decoupling.''  As the
universe expands, the momenta and temperature of neutrinos are simply
redshifted, and the neutrino temperature is given by the familiar
formulas
\begin{equation}
	T_\nu(a) = a^{-1}\,T_{\nu ,0}\,,\qquad
	T_{\nu ,0}=\left( {4\over 11}\right)^{1/3} T_{\gamma ,0}
		=1.947 K \,,
\end{equation}
where $a$ is the cosmic scale factor, the subscripts 0 denote the
present-day values, and the cosmic background photon temperature is
taken to be $T_{\gamma ,0}=2.728\,K$.~\cite{fix96}

An important feature of the neutrino distribution after decoupling is
that, although weak interactions are no longer rapid enough to keep
neutrinos in thermal equilibrium with other particle species,
neutrinos retain their equilibrium distribution as long as no other
physical processes (e.g., gravitational clustering; see Sec.~3) are
present to alter it.  Therefore, to zeroth order in density and metric
perturbations, the phase space distribution $f_0$ of the cosmic
background neutrinos is of the simple Fermi-Dirac form
\begin{equation}
	f_0(\epsilon)={g_s\over h_p^3} {1\over e^{\epsilon/
	k_B\,T_{\nu,0}}+1}\,,
\label{fermi}
\end{equation}
where $\epsilon=a(p^2+m_\nu^2)^{1/2}$ is the comoving energy,
$T_{\nu,0}$ is the neutrino temperature given by Eq.~(1), $g_s$ is the
number of spin degrees of freedom, and $h_p$ and $k_B$ are the Planck
and the Boltzmann constants.

The situation is further simplified if neutrino masses are $\ll 1$
MeV.  Such neutrinos are highly relativistic at decoupling; their
energy $\epsilon$, and hence the distribution function $f_0$, are
independent of $m_\nu$ to a good approximation.  One can easily show
that, as long as $m_\nu\ll 1$ MeV, the number density of the cosmic
background neutrinos is related to the neutrino temperature by
\begin{equation}
	n_\nu(T_\nu) = {7g_s\over 8\pi^2}\zeta(3) 
	\left( {k_B\,T_\nu\over \hbar\,c}\right)^3 \,,
\end{equation}
where $\zeta(3)\approx 1.202$ is the Riemann zeta function of order 3.
This gives a present-day density of $\approx 113$ cm$^{-3}$ for every
neutrino species independent of their masses.  (For comparison, the
present-day photon density is $\approx 412$ cm$^{-3}$.)  It also
follows that the contribution of these neutrinos to the present-day
mass density parameter, $\Omega_\nu$, is related to their masses by
the simple relation
\begin{equation}
        \Omega_\nu\,h^2 = {\Sigma_i m_i \over 93\, {\rm eV}}\,,
\end{equation}
where the index $i$ runs over all light, stable neutrino species
(e.g., $\nu_e, \nu_\mu$, and $\nu_\tau$), and the Hubble constant is
$H_0=100\,h$ km s$^{-1}$ Mpc$^{-1}$. One then arrives at the important
conclusion that in order for neutrinos not to close universe (i.e.
$\onu\le 1$), the sum of neutrino masses must not exceed $93\,h^2$ eV.
This value is far below the current laboratory limits (see Sec.~1).
Cowsik \& McClelland~\cite{cm72} were the first to use such
cosmological arguments to place an upper bound on neutrino masses.
(Unfortunately, these ``hot dark matter'' models in which the mass
density is dominated by massive neutrinos have been found to produce
excessive large voids surrounded by large coherent sheets and
filaments that are not seen in the observable universe.~\cite{white83}
Modifications to this model will be discussed below.)

In the high mass regime, $m_\nu \gg 1$ MeV, there exists another
window where the neutrino contribution to the mass density parameter
$\Omega$ of the universe is subcritical.  The argument is that
neutrinos with $m_\nu \gg 1$ MeV become non-relativistic long before
decoupling.  Neutrino and anti-neutrino pairs cease to be created in
abundance once the thermal temperature drops below $m_\nu$, and the
neutrino density is suppressed by the Boltzmann factor
$e^{-m_\nu/k_B\,T}$.  This large reduction factor in the relic
abundance allows neutrinos to have large masses without overclosing
the universe.  A more careful calculation~\cite{lee} shows that an
$\Omega\le 1$ universe implies a {\it lower} limit of $\sim 2$ GeV if
these heavy neutrinos are Dirac, and $\sim 6$ GeV if they are
Majorana.  Since this mass range is well above the current upper mass
bounds from laboratory measurements, it is of interest only when one
considers more exotic theories for neutrinos.~\cite{pgl}

\subsection{Kinematics and Free Streaming}

Let us now turn to the kinematics and the streaming properties of
neutrinos.  In general, neutrinos of mass $m _\nu$ become
non-relativistic after a redshift of
\begin{equation}
	z_{\rm rel} \approx {m_\nu c^2 \over 3k_B\,T_{\nu ,0}}
	= 2\times 10^3 \left( {m_\nu \over 1\, {\rm eV}} \right)  \,.
\end{equation}
This redshift has important implications for structure formation
because it dictates the time at which massive neutrinos begin to make
a transition from being radiation to matter.  Note that this
transition occurs fairly early, before recombination if $m_\nu\go 1$
eV.  The average momentum of the cosmic background neutrinos at
temperature $T_\nu$ is given by
\begin{equation}
        \left< p \right > = 3.15\,k_B\, T_\nu/ c\,.
\end{equation}
In the non-relativistic regime ($p=m_\nu\,v$), the average neutrino
speed can be written as
\begin{equation}
        \left< v \right> = 160\, {\rm km/s\ } \left( {1\, {\rm eV}\over
        m_\nu} \right) \left( {T_\nu\over 1.947} \right) \,.
\label{vave}
\end{equation}
Since $T_\nu \propto a^{-1}\propto (1+z)$, massive neutrinos slow down
as time goes on.  It is important to keep in mind that neutrinos with
a mass of several eV have slowed down to an average velocity below 100
km s$^{-1}$ today.

We also note that at the redshift of matter-radiation equality,
$z_{\rm eq} \sim 24000\,\Omega\,h^2$ (i.e. when the total energy
density in radiation in the universe equals that in matter), light
neutrinos with $1 < m_\nu < 10$ eV are zooming around with speeds
close to $c$.  Such large thermal speeds prevent massive neutrinos
from clustering gravitationally during this epoch, and this is why light
neutrinos are referred to as hot dark matter (HDM).  In contrast,
perturbations in cold dark matter (CDM), which by definition has
negligible thermal velocities, can grow unimpeded after $z_{\rm eq}$.
I will quantify the different clustering behavior of CDM and HDM
further in Sec.~3 and 4.

Since neutrinos cannot cluster appreciably via gravitational
instabilities on scales below the free streaming distance, this
introduces a characteristic length scale into the problem.  This scale
is given by the free-streaming wavenumber (in comoving coordinates)
\begin{equation}
	\kfs^2= {4\pi G\rho\, a^2\over \left< v \right>^2 }\,,
\end{equation}
which is analogous to the Jeans length for a self-gravitating system
of density $\rho$.  For $k<\kfs$ (i.e. large wavelengths) , the density
perturbation in the neutrinos is Jeans unstable and grows unimpeded in
the matter-dominated era.  For $k>\kfs$, the density perturbation
decays due to neutrino phase mixing.  A phase-space interpretation of
the free streaming property is that the phase mixing of collisionless
particles damps the growth of density perturbations.~\cite{bond83}
When the neutrinos are relativistic, $\left< v\right> \approx c\,$,
and the free-streaming distance is approximately the particle horizon,
which scales as $\kfs(a) \propto a^{-1}$ (in the radiation-dominated
era).  After the neutrinos become non-relativistic the relations
$\left< v\right> \propto a^{-1}m_\nu^{-1}$, $m_\nu\propto\onu h^2$,
and $\rho\propto a^{-3} h^2$ then imply~\cite{ma96}
\begin{equation}
	\kfs(a)\propto a^{1/2}\onu h^3\,.
\label{kfs2}
\end{equation}
As expected, the free-streaming distance ($\propto \kfs^{-1}$)
decreases with time as the neutrinos slow down.  This also implies
that neutrinos can cluster gravitationally on increasingly small
length scales at later times.~\cite{ma96} Such behavior has been seen
in cosmological numerical simulations and will be discussed in Sec.~4.

\section{Linear Perturbations in Neutrinos and Other Particles}
Thus far our discussion has focused on the properties of the smooth
cosmic background neutrinos, and Eqs.~(1)-(7) were derived under this
assumption.  The universe today, however, is clearly far from being
homogeneous on scales of $\sim 100$ Mpc and below.  Baryons and dark
matter in galaxies, clusters, and superclusters show a wide spectrum
of overdensities above the cosmic mean.  The current theoretical
framework for the origin and evolution of these cosmic structures
rests upon the assumption that certain primordial fluctuations
(perhaps originated from quantum fluctuations of scalar fields during
the inflationary era) imprint a perturbation spectrum on all matter
and radiation.  These fluctuations subsequently grow via gravitational
instabilities to give rise to the wide range of observed structures.
How are the cosmic relic neutrinos affected by all this?

To understand the growth of density perturbations in neutrinos as well
as other forms of matter and radiation, one would need to learn the
linear cosmological perturbation theory of gravitational instability.
A full description of this theory requires more time than is allocated
for these lectures.  I will only sketch the theory below with emphasis
on the neutrino component.  Interested readers should refer to the
pioneering work of Lifshitz, later reviewed in Lifshitz \&
Khalatnikov.~\cite{lifshitz} More modern treatments of various aspects
of this theory can be found in the textbooks by Weinberg and
Peebles,~\cite{text} in the reviews by Kodama \& Sasaki and Mukhanov,
Feldman \& Brandenberger,~\cite{review} and in the Summer School
lectures by Efstathiou, Bertschinger, and Bond.~\cite{school} A
complete description of this theory for all relevant particles is
given by Ma \& Bertschinger.~\cite{mb95} Here, I will only discuss the
essence of the theory and highlight the physical meaning of the key
results.

\subsection{Neutrino Phase Space}
Let us start with the neutrinos.  The full phase space distribution
function of neutrinos can be written as
\begin{equation}
	f(\vec{x},\vec{p},t)=f_0(p)+f_1(\vec{x},\vec{p},t)\,,
\end{equation}
where $f_1$ denotes perturbations to the Fermi-Dirac distribution
$f_0$ given by Eq.~(\ref{fermi}).  Unlike the unperturbed term $f_0$
that depends only on $p$, $f_1$ can have complicated dependence on
time as well as positions $\vec{x}$ and the conjugate momenta
$\vec{p}$.  The equations for neutrino temperature, number density
etc. discussed in Sec.~2 were obtained assuming $f=f_0$.  A
non-vanishing $f_1$ would lead to perturbations in these quantities.
For example, the perturbed neutrino energy density is related to $f_1$
by
\begin{equation}
  \delta\rho(\vec{x},t)= a^{-4}\int d^3 p\,\epsilon\,f_1(\vec{x},\vec{p},t) \,.
\end{equation}
Other quantities such as perturbations in the pressure and shear can
also be related to $f_1$.

It is in general difficult to compute and sample $f_1$ directly
because at a given time, it depends on six variables.  A Monte Carlo
technique, or a ``general-relativistic $N$-body'' technique, has been
developed to evolve $f_1$ from redshift $z\sim 10^9$ shortly after
neutrino decoupling until $z\sim 10$ when nonlinear effects become
non-negligible.~\cite{mb94a} In this calculation, an ensemble of
neutrino simulation particles is initially assigned velocities drawn
from the Fermi-Dirac distribution, which is an excellent approximation
at $z\sim 10^9$.  The trajectory of each neutrino simulation particle
is then followed by integrating the geodesic equations in the {\it
perturbed} background spacetime.  The metric perturbation gives rise
to a nonzero $f_1$, and the particle positions and velocities at a
later time $t$ represent a realization of $f_1(\vec{x},\vec{p},t)$.
Results from this calculation have revealed that at $z\sim 15$,
positive correlations have developed in the rms neutrino velocities
and the overdensity, which would be absent if the phase-space
distribution were purely Fermi-Dirac (i.e. $f=f_0$).  The more
spatially clustered neutrinos are found to move faster, possibly
resulting from an increase in the kinetic energy during gravitational
infalls.

\subsection{Evolution of Perturbed Density Fields}
The phase-space description above is applicable to all particle
species and is used in the full theory.  The full,
general-relativistic version of the linear perturbation theory is
described by a set of coupled and linearized Einstein, Boltzmann, and
fluid equations.  The variables include the metric perturbations to
the homogeneous and isotropic Friedmann-Robertson-Walker spacetime,
and the phase-space perturbations in all relevant particle species
(e.g., photons, baryons, cold dark matter, massless and massive
neutrinos).  The Einstein equations describe how the time evolution of
the metric perturbations is affected by the perturbations in the
density, pressure, shear, and higher-order moments of matter and
radiation.  The Boltzmann and fluid equations, on the other hand,
describe the time evolution of the radiation and matter distribution
in the perturbed spacetime.  Together, this theory describes the
growth of metric and density perturbations throughout the early
history of the universe, and it serves as the foundation for all
calculations of the linear power spectra for matter and
temperature variations imprinted on the cosmic microwave background.

\begin{figure}
\epsfxsize=5.truein 
\epsfbox{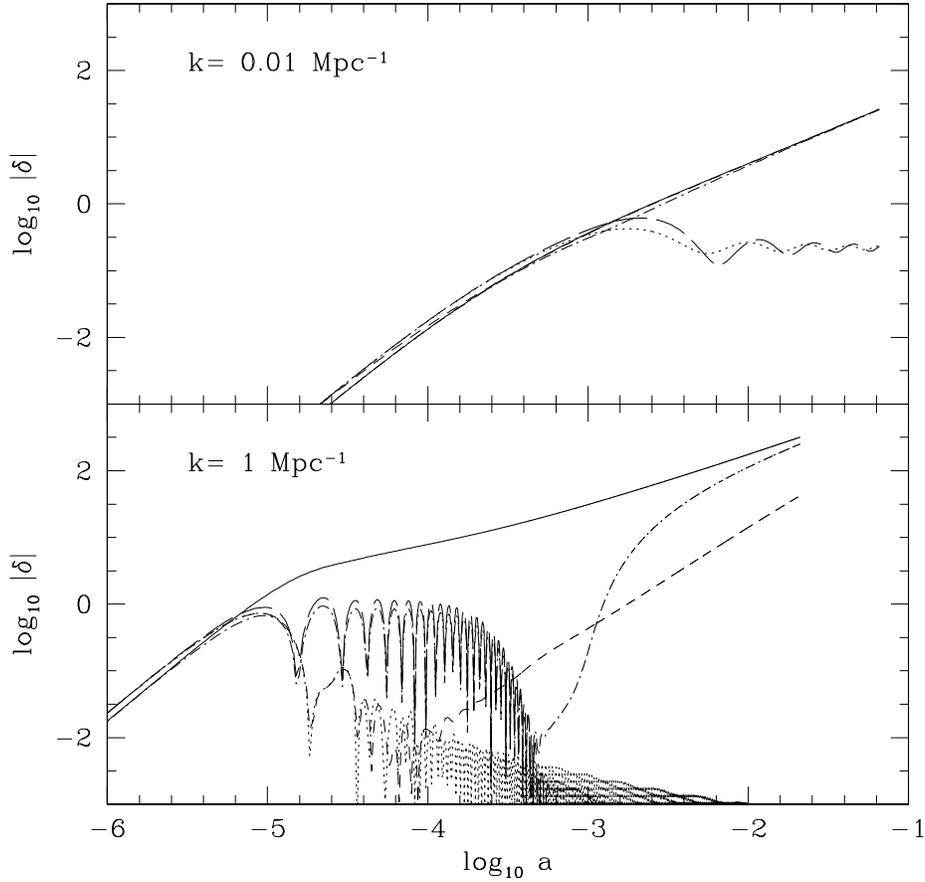}
\caption{Time evolution of the perturbed energy density field,
$\delta=\rho/\bar{\rho}-1$, for five matter and radiation components
in a flat C+HDM cosmological model.  (See text for model parameters.)
The results are from integration of the coupled Einstein and Boltzmann
equations.  Since the equations are linearized, each $k$-mode evolves
independently. Two modes are shown here for illustration.  In each
panel, the five curves represent $\delta$ for the cold dark matter
(solid), baryons (dash-dotted), photons (long-dashed), massless
neutrinos (dotted), and massive neutrinos (short-dashed),
respectively.}
\label{fig:del}
\end{figure}

Figure~\ref{fig:del} illustrates a small subset of results that can be
obtained from numerical integration of these linearized equations.  It
shows the time evolution of the perturbed energy density field,
$\delta=\delta\rho/\bar{\rho}=\rho/\bar{\rho}-1$, for the five
relevant particle species in a cold+hot dark matter (C+HDM) model.
This model assumes an Einstein-de Sitter universe containing a mixture
of CDM, HDM (i.e. massive neutrinos), baryons, photons, and massless
neutrinos.  The density parameters in the first three components in
this model are $\oc=0.75$, $\onu=0.2$, and $\ob=0.05$, and the Hubble
parameter is taken to be $H_0=50$ km s$^{-1}$ Mpc$^{-1}$, or $h=0.5$.
From Eq.~(4), these parameters correspond to a neutrino mass of 4.7
eV.  (For definiteness, this calculation has assumed that only one
type of neutrinos, presumably $\nu_\tau$, has a non-negligible mass.)
Two wavenumbers, $k=0.01$ Mpc$^{-1}$ (top) and 1 Mpc$^{-1}$ (bottom),
are shown in Figure~\ref{fig:del} to demonstrate the intricate
dependence of $\delta$ on length and time scales.  The overall
normalization of $\delta$ is set arbitrarily.

We observe several salient features in Figure~\ref{fig:del}.  First,
the amplitudes of $\delta$ for all particles grow monotonically until
a critical time, after which different particle species exhibit very
different behavior.  This critical time is the ``horizon crossing''
time, and it occurs when the horizon has grown large enough to
encompass the wavelength of a given mode of perturbation.  A mode of
perturbation is therefore not in causal contact until horizon
crossing.  Naturally, this occurs earlier for smaller wavelengths
(i.e. larger $k$).  In Figure~\ref{fig:del}, one can indeed see that
the $k=1$ Mpc$^{-1}$ mode enters the horizon at $a\sim 10^{-5}$ while
the $k=0.01$ Mpc$^{-1}$ mode enters the horizon a little after $a=
10^{-3}$.  A point to keep in mind is that the behavior of $\delta$
before horizon crossing is strongly dependent on the choice of gauge.
The results shown in Figure~\ref{fig:del} are computed in the
so-called synchronous gauge, which is a popular choice due to
historical precedent.~\cite{lifshitz} See Mukhanov, Feldman \&
Brandenberger~\cite{review} and Ma \& Bertschinger~\cite{mb95} for
discussion of a more convenient gauge (the conformal Newtonian gauge).

The second feature in Figure~\ref{fig:del} to note is that after
horizon crossing, the photons (long-dashed) and baryons (dot-dashed)
exhibit rapid, coupled oscillations in the $k=1$ Mpc$^{-1}$ mode but
only the photons oscillate in the $k=0.01$ Mpc$^{-1}$ mode.  This
occurs because the former enters the horizon before recombination at
$a_{\rm rec} \sim 10^{-3}$, and the photons and baryons are coupled by
Thomson scattering and oscillate acoustically.  The coupling is not
perfect.  The friction of the photons dragging against the baryons
leads to Silk damping,~\cite{silk68} which is prominent in the bottom
panel of Figure~\ref{fig:del} at $a\sim 10^{-3.5}$.  After
recombination, the baryons decouple from the photons and fall quickly
into the potential wells formed earlier by the CDM.  This results in
the rapid growth of the dot-dashed curve in the bottom panel of
Figures~\ref{fig:del}.  The mode with $k=0.01$ Mpc$^{-1}$, on the
other hand, enters the horizon when the universe is neutral.  The
baryons therefore grow like the CDM and do not oscillate.  The
critical length scale demarcating these two regimes is the horizon
size at recombination $a_{\rm rec}$: $k_{\rm rec} \sim 0.03$
Mpc$^{-1}$.

The third feature to note in Figure~\ref{fig:del} is the rate of
growth of the CDM component (solid curve) after horizon crossing.
Close inspection shows that the CDM in the bottom panel grows more
slowly at $10^{-5}\lo a \lo 10^{-4}$ than later on, whereas in the
upper panel, the CDM simply grows with a power law after horizon
crossing.  This is because the shorter wavelength mode ($k=1$
Mpc$^{-1}$) enters the horizon when the energy density of the universe
is dominated by radiation, and fluctuations in matter (e.g. CDM)
cannot grow appreciably during this era.  The critical scale
separating continual and suppressed growth is the horizon size at the
time of radiation-matter equality $a_{\rm eq} \sim 4\times 10^{-5}
(\Omega h^2)^{-1}$: $k_{\rm eq} \sim 0.1$ Mpc$^{-1}$ for the
parameters of this model.

The fourth feature to note in Figure~\ref{fig:del} is the behavior of
the massive neutrinos (short-dashed).  As discussed in Sec.~2,
neutrinos of masses within the cosmologically interesting range ($\sim
1$ to 10 eV) are highly non-relativistic today but were relativistic
at earlier times.  This property is in fact evident in the top panel
of Figure~\ref{fig:del}: Careful inspection shows that at $a\approx
10^{-4}$, the density field $\delta$ in massive neutrinos is indeed
making a gradual transition from the upper line for the radiation
fields to the lower line for the matter fields.  (More precisely, the
primordial perturbations are assumed to be ``isentropic'' here, which
leads to a perturbation amplitude that is a factor of $4/3$ higher for
radiation than matter.)  The subsequent evolution of $\delta$ in
massive neutrinos for this mode ($k=0.01$ Mpc$^{-1}$) is very similar
to that of CDM.  This is because it enters the horizon when the
thermal velocities of the neutrinos have decreased substantially; the
free-streaming effect is therefore unimportant.  For the $k=1$
Mpc$^{-1}$ mode, on the other hand, the free streaming effect is
evident and the growth of $\delta$ in massive neutrinos is suppressed
until the characteristic free-streaming scale $k_{\rm fs}(a)$ given by
Eq.~(9) grows to $\sim k$.  Afterwards, the short-dashed curve for
massive neutrinos is seen to grow again and catch up to the CDM.
Since $k_{\rm fs} \propto a^{1/2}$, the larger $k$ modes suffer more
free-streaming damping and $\delta$ for massive neutrinos can not grow
until later times.  The damping in the massive neutrino component also
affects the growth of the CDM, slowing it down more for models with
larger $\onu$ compared to the pure CDM model.

\subsection{Growth Rate and Power Spectrum}
I have used Figure~\ref{fig:del} computed for two particular $k$-modes
in a particular cosmological model to illustrate the physical meaning
of many key features in the evolution of the density field for matter
and radiation throughout the cosmic history.  I will now discuss
general descriptions that can be conveniently used to characterize the
fluctuation amplitudes over a range of length scales for a variety of
cosmological models.  For example, it is extremely useful to know the
dependence of $\delta$ and its time derivative on $k$ at a given time
for a wide range of models.  The most basic quantity to use is the
linear power spectrum, $P(k,t)$, and the growth rate of the density
field, $f\equiv d\log\delta/d\log a$. ({\em Note: I am following the
convention of using $f$ to denote the growth rate; it should not be
confused with the phase-space distribution function of Sec.~2 and
3.1.})  The power spectrum quantifies the two-point statistics of
$\delta$, and for a Gaussian field, $P(k)$ represents its rms
fluctuations and completely specifies its statistical properties.  The
power spectrum is therefore of fundamental importance in cosmology.

\begin{figure}
\epsfxsize=3.7truein 
\epsfbox[0 244 482 653]{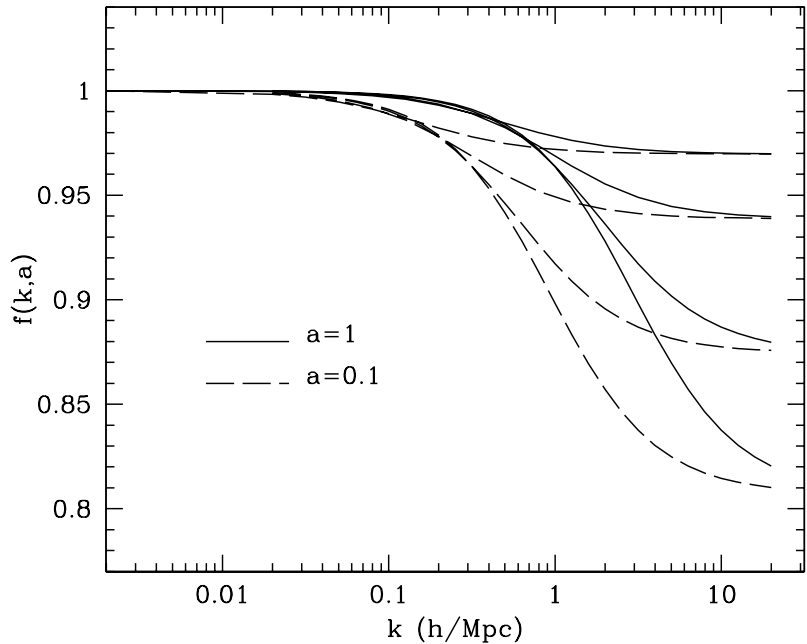}
\caption{Growth rate of the CDM density field, $f\equiv
d\log\delta/d\log a$, in four flat C+HDM models at $a=1$ (solid) and
0.1 (dashed).  The four models assume different neutrino masses:
$m_\nu=1.2$, 2.3, 4.6, and 6.9 eV (from top down), corresponding to
$\onu=0.05$, 0.1, 0.2, and 0.3.  At small $k$, the CDM density field
in these models grows with the same rate ($\delta\propto a$) as in the
standard CDM model.  At large $k$, the growth rate is suppressed
because a fraction of the energy density in C+HDM models is in the hot
neutrinos that exhibit less gravitational clustering.  The suppression
at large $k$ becomes less severe at later times because the velocities
of the hot neutrinos decrease with time.}
\label{fig:f}
\end{figure}

Comparing the growth rate and power spectrum for models with and
without massive neutrinos is an effective way to illustrate the
effects of hot dark matter.  In the standard CDM model with $\Omega=1$
and $h=0.5$ (neutrino mass is assumed to be zero), the CDM density
field grows as the expansion factor $a$ on all scales; therefore
$f=1$.  As discussed in Sec.~2.2, massive neutrinos introduce an
additional length scale, the free-streaming distance, below which
fluctuations are washed out and the growth rate is retarded.  The
growth rate is therefore more intricate in models with massive
neutrinos and is generally a function of the wavenumber $k$, neutrino
density parameter $\onu$, and time.  Figure~\ref{fig:f} illustrates
such dependence in four different C+HDM models that assume a mixture
of CDM and HDM.  It shows that the growth is suppressed at large $k$,
and models with a larger fraction of energy density in HDM suffer
more.  It also shows that the suppression becomes less severe at later
times.

One can gain some understanding of the behavior shown in
Figure~\ref{fig:f} by exploring two asymptotic regions that can be
solved analytically for C+HDM models with $\oc+\onu=1$: (1) Since HDM
behaves like CDM above the free-streaming distance, $f\rightarrow 1$
as $k\rightarrow 0$; (2) In the opposite limit of large $k$, the HDM
density field $\delta_h$ is severely dampened compared to the CDM
density field $\delta_c$ because of the neutrino effects.  For
$\delta_h \ll \delta_c$, the time evolution of the CDM density field
is governed by the linearized fluid equation
\begin{equation}
  \ddot{\delta_c}+{\dot{a}\over a}\dot{\delta}_c=1.5 H^2 a^2\oc \delta_c\,,
\end{equation} 
where the dots denote differentiation with respect to the conformal
time $\tau$.  Since $Ha=2/\tau$ in the matter-dominated era, the
growing solution in this regime is easily shown to be~\cite{bond80}
\begin{equation}
        f_\infty\equiv f(k\rightarrow\infty)
        ={1\over 4}\sqrt{1+24\oc}-{1\over 4}
        ={5\over 4}\sqrt{1-{24\over 25}\onu}-{1\over 4}\,.
\label{finfy} 
\end{equation} 
It is interesting to note that
\begin{equation}
        f_\infty \approx \oc^{0.6} 
\end{equation} 
is an excellent approximation to the equation above, especially for
the cosmologically viable range of $\onu\lo 0.3$.  Using these
analytic solutions in the asymptotic regimes and the scaling
dependence of the free streaming wavenumber $\kfs$ in
Eq.~(\ref{kfs2}), one can construct a simple approximation for $f$ for
a wide range of model parameters.  It is found that the growth rate
$f$ is well approximated by~\cite{ma96}
\begin{equation}
    f \equiv {d\log \delta\over d\log a} = {1+ \oc^{0.6}\,
    0.00681\,x^{1.363} \over 1+0.00681\,x^{1.363}}\,,\qquad x\equiv
    {k\over \Gamma_\nu h} \,,
\label{f}
\end{equation}
where $\Gamma_\nu$ is a shape parameter derived from Eq.~(\ref{kfs2}),
\begin{equation}
	\Gamma_\nu=a^{1/2}\onu h^2\,,
\end{equation}
$\oc+\onu=1$, and $k$ is in units of Mpc$^{-1}$.  Note that
Eq.~(\ref{f}) depends only on the variable $x$ that characterizes the
neutrino free-streaming scale, and $\onu$ (or $\oc$) via $f_\infty$.
The fractional error of the fit relative to the numerically computed
values is smaller than 0.5\% for a wide range of parameters.  The
seemingly complicated multi-parameter dependence of Figure~\ref{fig:f}
is succinctly incorporated in Eq.~(\ref{f}).

\begin{figure}
\epsfxsize=3.6truein 
\epsfbox[0 130 482 653]{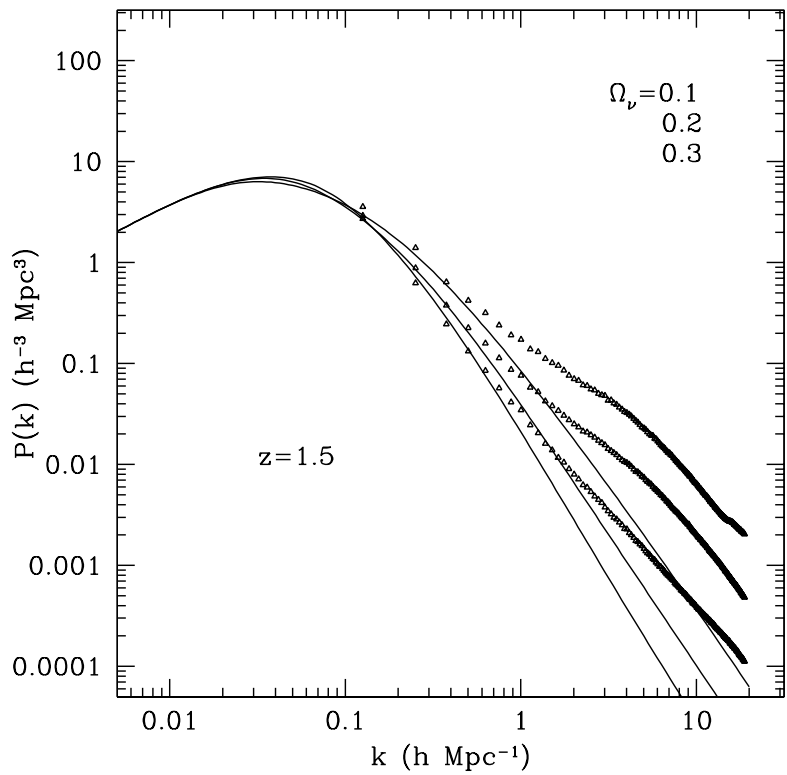}
\caption{Power spectrum at $z=1.5$ for the perturbed matter density
field for three flat C+HDM models with $\onu=0.1$, 0.2, and 0.3.  The
solid curves show the linear $P(k)$ computed from the linear
perturbation theory; accurate analytical fitting formulas are given by
Eq.~(\ref{g}).  The triangles show the nonlinear $P(k)$
computed from $N$-body simulations; accurate analytical approximations
are given by Eq.~(\ref{master}).}
\label{fig:pow}
\end{figure}

For the linear power spectrum, the slower time growth of $\delta$ at
$k > \kfs$ for the C+HDM models shown in Figure~\ref{fig:f} indicates
a suppressed clustering amplitude on these scales.  This effect is
illustrated in Figure~\ref{fig:pow}, which shows the density-averaged
power spectrum (i.e. $P=\{\oc \sqrt{P_c}+\onu \sqrt{P_\nu}\}^2$) for
three flat C+HDM models.  In general, $P_\nu \ll P_c$ on small length
scales due to the neutrino thermal velocities, and the models with
higher $\onu$ clearly have less power at large $k$ in accordance with
Figure~\ref{fig:f}.  A good approximation for the relative $P(k)$ in a
flat C+HDM model (with $\onu\lo 0.3$) and a pure CDM model $(\onu=0$) is
given by~\cite{ma96}
\begin{eqnarray}
  && {P(k,a,\onu)\over P(k,a,\onu=0)}
        = \left( { 1+b_1\,x^{b_4/2}+b_2\,x^{b_4} \over 1+b_3\,x_0^{b_4} }
        \right)^{\onu^{1.05}}\,,\nonumber\\
   && x={k\over \Gamma_\nu} \,, \quad x_0=x(a=1) \,,
\label{g}
\end{eqnarray} 
where the best-fit parameters are $b_1=0.004321, b_2=2.217\times
10^{-6}, b_3=11.63$, and $b_4=3.317$ for $k$ in units of Mpc$^{-1}$.
Analytical approximations for the separate cold and hot spectra $P_c$
and $P_\nu$ can be found in the same reference.  More complicated
approximations for a wider range of parameter space have also been
proposed.~\cite{eh98}

\section{Nonlinear Gravitational Clustering of Neutrinos}
As far as cosmological structure formation is concerned, the main
difference between massive and massless neutrinos is that the former
can participate in the processes of gravitational clustering and hence
serves as a component of the dark matter in the universe.  Massless
neutrinos, on the other hand, affect cosmology only through their
contribution to the radiation energy density.  Any primordial
perturbations in massless neutrinos are damped out after horizon
crossing as a result of phase mixing (see dotted curves in
Figure~\ref{fig:del}), and the only remnant of this component is the
elusive $T_{\nu ,0}=1.947$ K background described in Sec.~2.

\subsection{Spatial Distribution of Neutrinos}
For massive neutrinos in models with a mixture of CDM and HDM, it is
interesting to ask: do the neutrinos fall in the CDM potential wells
and form a part of dark matter halos?  One may naively think not
because neutrinos are too hot.  Our discussion thus far, however,
indicates otherwise.  We have seen in Eq.~(\ref{vave}) that neutrinos
slow down with the expansion of the universe.  Those with a mass of
several eV are travelling with a speed much below the typical velocity
dispersions $\sim 200$ km s$^{-1}$ of stars in galaxies.  They can
potentially be bound to galactic halos.  In general, the extent to
which massive neutrinos can cluster gravitationally depends on their
mass and speed.

\begin{figure}
\epsfxsize=5.truein 
\epsfbox[80 144 592 718]{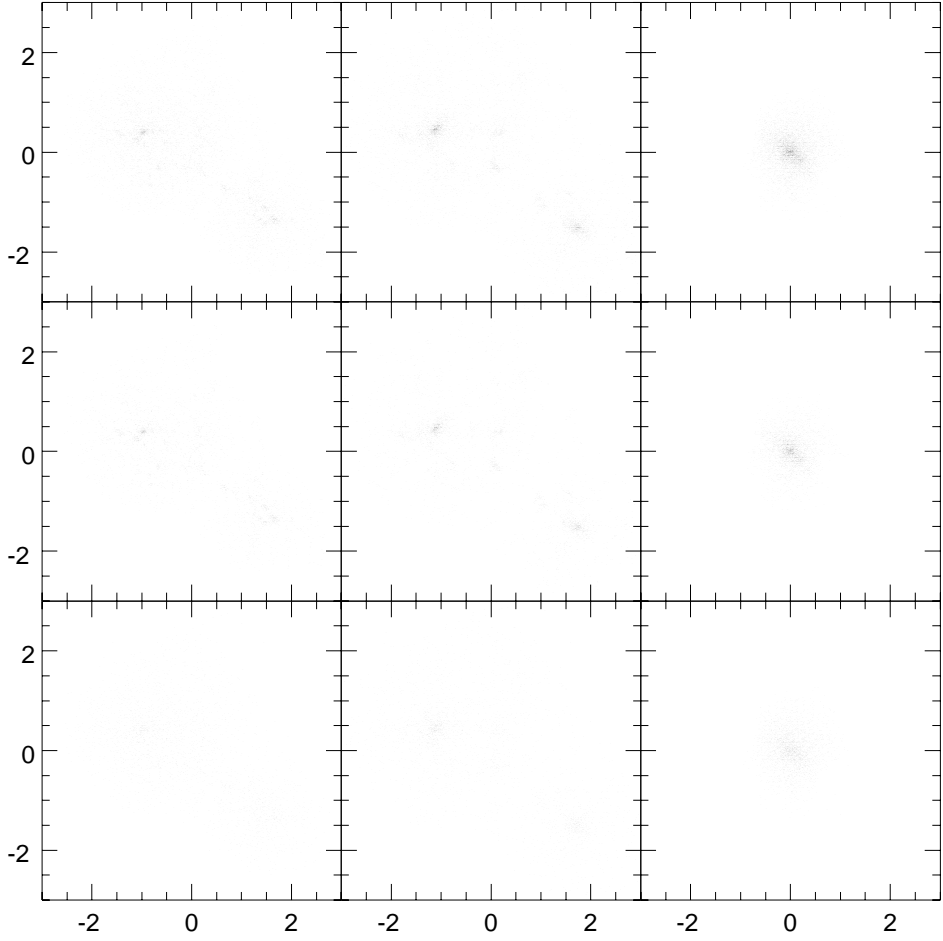}
\caption{Spatial distribution of cold dark matter (middle) vs. hot
dark matter (bottom) particles in a simulated halo formed in a large
$N-$body run for the flat $\onu=0.2$ C+HDM model.  The top panels
shows the sum of cold and hot particles.  Three redshifts are shown:
$z=2$, 1, and 0 (from left to right), corresponding to cosmic times of
2.5, 4.6, and 13 billion years.  All boxes have the same physical
scale, $3.5\times 3.5\,h^{-1}$ Mpc.  The final halo at $z=0$ is
clearly a merger product of two dominant subhalos and several smaller
satellites at higher redshifts.  The spatial distribution of the
massive neutrinos (4.7 eV) is visibly smoother than that of the CDM.}
\label{fig:halo}
\end{figure}

To illustrate this point further, let us examine results from
numerical simulations.  Figure~\ref{fig:halo} shows the projected
spatial distribution of cold particles (middle), hot particles
(bottom), and the sum of the two (top) in a simulated dark matter halo
in a flat $\onu=0.2$ C+HDM model ($m_\nu=4.7$ eV).  Three redshifts
are shown (from left to right): $z=2, 1$ and 0 when the universe is
2.5, 4.6, and 13 billion years old, respectively.  Each panel is
$3.5\times 3.5\,h^{-1}$ Mpc in {\it physical} coordinates.  The parent
simulation is a large $N$-body run with 23 million simulation
particles in a (100 Mpc)$^3$ comoving box.~\cite{mb94b} The dense halo
shown at $z=0$ is clearly formed from mergers of two smaller halos and
their satellites at higher redshifts, demonstrating the ``bottom-up''
hierarchical pattern of structure formation which is preserved in
C+HDM models with $\oc > \onu$.  Massive neutrinos are visibly
clustered in the bottom panels, but their spatial distribution is
smoother than that of the CDM.  In the smaller CDM clumps at $z\go 1$,
there are no discernible HDM halos at the same locations.  The average
thermal speed of an ensemble of 4.7 eV Fermi-Dirac neutrinos at $z\go
1$ is $\go 70$ km/s (cf. Eq.~\ref{vave}).  As discussed in Sec.~3.1,
when linear perturbations are taken into account, neutrinos in
overdense regions have even higher velocities.  It is therefore not
surprising that the shallower potential wells of these small halos
cannot trap a substantial number of HDM particles.

The clustering of the 4.7 eV neutrinos shown in Figure~\ref{fig:halo}
should be compared with the constraint derived by Tremaine \&
Gunn,~\cite{tg79} which states that if cosmic neutrinos were to make
up the bulk of galactic and cluster halos, they must be more massive
than $\sim 10$ eV so that the Pauli exclusion principle is not
violated.  The C+HDM models considered here assume $\oc > \onu$ so as
to preserve the successful hierarchical formation of structure in pure
CDM models.  Dark matter halos such as in Figure~\ref{fig:halo} have a
substantial fraction of CDM, which in turn enhances the gravitational
infalls of massive neutrinos.  The clustering of neutrinos in C+HDM
models is therefore more complicated.

\subsection{Nonlinear Power Spectrum}
The process of nonlinear gravitational clustering can be quantified
statistically.  Here I will only discuss the lowest-order statistical
description given by the nonlinear power spectrum $P(k)$ of the matter
density field.

We have already discussed the linear power spectrum in Sec.~3.3.  The
triangles in Figure~\ref{fig:pow} show the nonlinear $P(k)$ computed
from the particle positions in numerical simulations of three
C+HDM models.  The hierarchical nature of gravitational collapse in
these models is illustrated by the fact that the high-$k$ modes have
become strongly nonlinear whereas the low-$k$ modes are still
following the linear power spectrum.  The fact that the lowest several
$k$ modes are still linear at $z=0$ ensures that our choice of the
simulation box size (100 Mpc) is large enough to include all waves
that have become nonlinear at present.

The calculation of the fully evolved $P(k)$ for a given model is a
laborious task involving the execution of high-resolution $N$-body
simulations.  Fortunately, some recent progress has been made in
constructing analytical fitting formulas for a wide range of
interesting models.  The strategy is to examine the mapping between
the linear and nonlinear $P(k)$ for a small, selected set of models
with $N$-body data and then to extract systematic behavior for a wider
range of parameters.  The work carried out thus far has investigated
scale-free models with a power-law power spectrum,~\cite{ham91,jain95}
pure CDM and CDM with a cosmological constant $\Lambda$ (LCDM)
models,~\cite{jain95,pd96,ma98} and C+HDM models.~\cite{ma98}  The
proposed formulas typically have a functional form that is motivated
by analytical solutions in asymptotic regimes, but in order to obtain
accurate approximations, the coefficients are calculated from fits to
the nonlinear $P(k)$ computed from the numerical simulations.
These approximations have provided physical insight into the process
of nonlinear collapse and much practical convenience in incorporating
the prominent nonlinear effects illustrated in Figure~\ref{fig:pow}.

The formula applicable for the widest range of cosmological models
thus far is given below.  It maps the density variance
$\Delta(k)\equiv 4\pi k^3P(k)$ in the linear and nonlinear regimes
by~\cite{ma98}
\begin{eqnarray}
       && {\Dnl(k)\over \Dl(k_0)} = G\left({\Dl(k_0) \over
        g_0^{1.5}\,\sigma_8^\beta} \right) \,,\nonumber\\ &&
        G(x)=[1+\ln(1+0.5\,x)]\,{1+0.02\,x^4 + c_1\,x^8/g^3 \over
        1+c_2\,x^{7.5}}\,.
\label{master}
\end{eqnarray}
Note that $\Dl$ and $\Dnl$ are evaluated at different wavenumbers,
where $k_0=k\,(1+\Dnl)^{-1/3}$ corresponds to the pre-collapsed length
scale of $k$.  The parameter $\sigma_8$ is the rms linear mass
fluctuation on $8\,h^{-1}$ Mpc scale at the redshift of interest, and
$\beta=0.7+10\,\onu^2$.  The functions $g_0=g(\om,\ov)$ and
$g=g(\om(a),\ov(a))$ are, respectively, the relative growth factor for
the linear density field at present day and at $a$ for a model with a
present-day matter density $\om$ and a cosmological constant $\ov$.  A
good approximation is given by~\cite{lahav}
\begin{equation}
	g ={5\over 2}\om(a) [
      \om(a)^{4/7}-\ov(a)+\left(1+ \om(a)/2\right) \left(1+ \ov(a)/70
      \right)]^{-1}\,,
\end{equation}
and $\om(a)=\om\,a^{-1}/[1+\om(a^{-1}-1) +\ov(a^2-1)]$ and
$\ov(a)=\ov\,a^2/[1+\om(a^{-1}-1)+\ov(a^2-1)]$.  The time dependence
is in factors $\sigma_8^\beta$ and $g$.  For CDM and LCDM models, a
good fit is given by $c_1=1.08\times 10^{-4}$ and $c_2=2.10\times
10^{-5}$.  For C+HDM, a good fit is given by $c_1=3.16\times 10^{-3}$
and $c_2=3.49\times 10^{-4}$ for $\onu=0.1$, and $c_1=6.96\times
10^{-3}$ and $c_2=4.39\times 10^{-4}$ for $\onu=0.2$.

\subsection{High-Redshift Constraints}

The C+HDM models discussed thus far are a class of models bridging
the much-studied albeit troubled pure CDM and pure HDM models.
They are parameterized by the neutrino density parameter $0<\onu< 1$,
or equivalently, by the neutrino mass $0< m_\nu <93 h^2$ eV (see
Eq.~(4)).  The original motivation for examining these mixed models is
to study whether the free-streaming effect introduced by the massive
neutrinos could suppress the growth of density perturbations below the
free-streaming scale, and thereby alleviate the problem of excess
small-scale clustering in the standard CDM model.  As illustrated in
Figure~\ref{fig:pow}, massive neutrinos do indeed reduce the amplitude
of clustering at large $k$, and a larger $\onu$ leads to smaller
high$-k$ power.

Any viable cosmological model that provides a good statistical match
to the local universe must also reproduce the appropriate evolutionary
history out to high redshifts.  Although gravitational clustering on
galactic scales is indeed reduced in C+HDM models, providing a better
match to low-redshift observations,\cite{ma96,chdm} this suppression in the
clustering power at high redshifts has been found to pose serious
problems for some models.  Studies based on semi-analytic theories and
dissipationless numerical simulations have shown~\cite{mb94b,dla} that flat
C+HDM models with $\onu>0.2$ do not produce enough early structure to
explain the statistics of damped Ly$\alpha$ systems at redshift $z\ge
2$.  More recent work~\cite{ma97} has included the effects of gas
ionization and dissipation in the theoretical calculations and has
compared the results to new data for damped Ly$\alpha$ systems at even
higher redshift $z\sim 4$.  It is found that ionization of hydrogen in
the outskirts of halos and gaseous dissipation near the halo centers
tend to exacerbate the problem of late galaxy formation.  The amount
of dense gas associated with the damped systems falls well below that
observed, even for the flat $\onu=0.2$ C+HDM model.  This has placed
an upper bound of $\sim 5$ eV on the sum of $\nu_e$, $\nu_\mu$, and
$\nu_\tau$ masses, which is much more stringent than the upper limits
given by current particle experiments (see Sec.~1).

Let us briefly discuss the limitations and uncertainties in these
calculations.  Still debated is the nature of damped Ly$\alpha$
absorption -- whether it is due to intervening large, rapidly-rotating
disk galaxies~\cite{pw97} with circular velocities $\go 200$ km
s$^{-1}$, or infalling and randomly moving protogalactic gas clumps in
dark matter halos~\cite{haeh98} with virial velocities of $\sim 100$
km s$^{-1}$.  When kinematic considerations are included, the latter
model may have problems balancing the high energy dissipation rate
caused by cloud collisions.~\cite{mm98} This uncertainty aside, in
either scenario, a host dark halo of velocity at least $\sim 100$ km
s$^{-1}$ is needed to reproduce the large velocity widths and
asymmetries of the observed low-ionization lines associated with
Ly$\alpha$ systems.  Uncertainties associated with the finite
resolution of simulations have also been studied in some
detail.~\cite{gardner97} It is found that even when the contribution
from the numerically unresolved halos with velocities $v \lo 100$ km
s$^{-1}$ is included, the absorption incidence is increased by at most
a factor of 2.  This is insufficient to erase the discrepancies
reported for flat C+HDM models with $\onu > 0.2$.  The upper bound on
neutrino masses from current cosmological studies is therefore $\sim
5$ eV.

\section*{Acknowledgments}

I thank the organizers P. Langacker and K. T. Mahanthappa of the
TASI-98 School for their hospitality.  This work was supported by the
National Scalable Cluster Project at the University of Pennsylvania,
the National Center for Supercomputing Applications, and a Penn
Research Foundation Award.

\end{document}